\documentclass[aps,preprint]{revtex4}

\usepackage{epsfig,amsmath}
\usepackage{subfigure}
\usepackage{graphicx}% Include figure files
\usepackage{dcolumn}% Align table columns on decimal point
\usepackage{stmaryrd}
\usepackage{mathrsfs}
\usepackage{pifont}
\usepackage{amsthm}
\usepackage{amssymb}
\usepackage{bm}
\usepackage{latexsym}
\usepackage{hyperref}
\usepackage{color}

\newcommand{\la}{\langle}
\newcommand{\ra}{\rangle}

\newcommand{\pa}{\partial}

\newcommand{\red}[1]{\textcolor{black}{#1}}

\def\pra#1{{ Phys.\ Rev. A\/} {\bf#1}}

\def\prl#1{{ Phys.\ Rev.\ Lett.} {\bf#1}}

\def\pla#1{{ Phys.\ Lett. A\/} {\bf#1}}

\begin{document}

\title{Universal Transformability between Hamiltonians and Hidden Adiabaticity}

\title{Hamiltonian Transformability, Fast Adiabatic Dynamics and Hidden Adiabaticity}

\author{Lian-Ao Wu$^{1,2}$\footnote{Author to whom any correspondence should be addressed. Email address: lianao.wu@ehu.es }, Dvira Segal $^{3,4}$ }

\affiliation{$^{1}$Department of Theoretical Physics and History of Science, The Basque Country University (EHU/UPV), PO Box 644, 48080 Bilbao, Spain \\ $^{2}$Ikerbasque, Basque Foundation for Science, 48011 Bilbao\\ 
$^{3}$Chemical Physics Theory Group, Department of Chemistry,
and Centre for Quantum Information and Quantum Control,
University of Toronto, 80 Saint George St., Toronto, Ontario, Canada M5S 3H6 \\
$^{4}$Department of Physics, University of Toronto, Toronto, Ontario, Canada M5S 1A7}
\date{\today}

%%%%%%%%%%%%%%%%%%%%
\begin{abstract}
We prove the  existence of a unitary transformation that enables two \red{arbitrarily given} Hamiltonians in the same Hilbert 
space to be transformed into one another. 
The result is straightforward yet, for example, 
it lays the foundation to implementing or mimicking dynamics 
with the most controllable Hamiltonian.
As a promising application, this existence theorem allows
for a rapidly evolving realization of adiabatic quantum computation
by transforming a Hamiltonian where dynamics is in the adiabatic regime into a rapidly evolving one. \red{We illustrate the theorem with examples.}

\end{abstract}


\maketitle

%===============================
%\section{Introduction}

{\em Introduction.--} Understanding quantum dynamics and control is essential 
to modern quantum technologies such as adiabatic quantum computation~\cite{AQC,Childs}. 
A quantum dynamical processes is driven by its corresponding Hamiltonian, where
the Hamiltonian represents a physical realization. For instance, spin dynamics can be driven by the 
Zeeman Hamiltonian, which is physically realized by applying magnetic fields~\cite{Messiah}.
Different realized dynamics, for example fast vs. adiabatically controlled passage~\cite{Rice1,Rice2, Berry09} may seem remote from one another.
However, in this paper we show that they \red{can} well be intimately related.

For example, a physical realization of adiabatic quantum computation 
(AQC) suffers from its slowness, with the resultant destructive effects of decoherence
and the occurrence of quantum phase transitions during dynamics~\cite{Adolfo,Torrontegui13,Jing13,JW2}.
Here, we \red{prove rigorously} that different dynamics, described by two Hamiltonians 
 defined on the same Hilbert space,
can always be transformed into one another.  
As a consequence, the physical outcome
of AQC can be made equivalent to the outcome of a dynamical process that 
can be extremely fast. 
This relationship between different dynamics
is based on a straightforward but profound proposition described below, 
implying, for example, that an adiabatic process
may be physically realized with a fast Hamiltonian. Similarly, it implies
a hidden adiabaticity amongst rapid dynamics.

%==============================================================
%\section{The Transformability proposition}\label{theory}

{\em The Transformability proposition.--} 
Given any two Hamiltonians, $\hat H$ and $\hat h$ in the same Hilbert space, which can be time-independent 
or time-dependent, the corresponding Schr\"odinger equations are
\begin{equation}\label{e1}
i\pa_t\hat U=\hat H(t)\hat U,
\end{equation}
and
\begin{equation}\label{e2}
i\pa_t\hat u=\hat h(t)\hat u,
\end{equation}
where $\hat U$ and $\hat u$ are propagators of $\hat H(t)$ and $\hat h(t)$, respectively. 
%We now prove the following proposition. 

Proposition: {\em  Two Hamiltonians $\hat H$ and $\hat h$ can always be transformed into
one another. }
Mathematically, this claim can be expressed as: For given $\hat H$ and $\hat h$, 
there exists at least  one unitary operator $\hat S$ such that 
\begin{equation}\label{e3}
\hat h=\hat S^{\dagger} \hat H \hat S-i\hat S^{\dagger}\dot{\hat S},
\end{equation}
and
\begin{equation}\label{e4}
\hat H=\hat S \hat h \hat S^{\dagger}-i\hat S\dot{\hat S}^{\dagger}.
\end{equation}
where the overdot indicates a time derivative.
Proof: The operator $\hat S$ enables the transformation $\hat U=\hat S\hat u$. 
Substituting it into the Schr\"odinger equation~(\ref{e1}), we obtain Eq.~(\ref{e2}) with 
the {\em effective} Hamiltonian $\hat h=\hat S^{\dagger} \hat H \hat S-i\hat S^{\dagger}\dot{\hat S}$. 
Similarly,   if we begin with the Schr\"odinger equation (\ref{e2})  we
transform it to Eq. (\ref{e1}) by identifying its Hamiltonian with
$\hat H=\hat S \hat h \hat S^{\dagger}-i\hat S\dot{\hat S}^{\dagger}$.
Because the solutions $\hat u$ and $\hat U$ of the Schr\"odinger equations~(\ref{e1}) and (\ref{e2}) always exist,
 so does the product $\hat U\hat u^{\dagger}$. 
\red{By setting} $\hat S=\hat U\hat u^{\dagger}$, we can reproduce the Hamiltonians~(\ref{e3}) and (\ref{e4}) and 
therefore formally prove the \red{universal} existence of the \red{unitary} transformation $\hat S$. \red{In other words, there is {\em always} a unitary transformation that enables two arbitrarily given Hamiltoni- ans in the same Hilbert space to be transformed into one another.}
We term this property transformability, and the two Hamiltonians $\hat H$ and $\hat h$ are {\em transformable}.
The special case of the proposition with $\hat h$ being time-independent was proven a quarter-century ago in 
Ref.~\cite{PLA93}.

%=============================================
{\em Rapid Adiabatic Quantum Computation}

Adiabatic quantum computation is one of the most promising candidates 
to realize quantum computing \cite{Lidar}. The approach is based on the adiabatic theorem: 
The solution to a computational problem of interest is encoded in the ground state of a 
potentially complicated Hamiltonian. To approach the solution, one prepares
a system with a simpler Hamiltonian and initializes it at its ground state.
By evolving the Hamiltonian sufficiently slowly towards the desired (complex) one, 
the adiabatic theorem guarantees that the system
follows the instantaneous ground state, finally realizing the target ground state.
Evidently, the slowness of AQC could be the main impediment to its utility for 
quantum algorithms.

The universal transformability property suggests that a slow AQC process $\hat u$ 
can be mapped onto a fast quantum process $\hat U$---that is more controllable, and suffers
reduced decoherence during processing. 
Note that as a convention, we use lower (upper) cases to denote the slow (fast) dynamics throughout the paper.]
Consequently, AQC can be physically realized by a fast process.

The eigenstate  $|E(T)\ra$ of the problem Hamiltonian at time $T$ is given by implementing the 
adiabatic process
\begin{equation}
|E(T)\ra \sim  \hat u|E(0)\ra=\hat S^{\dagger} (T) \hat U (T) |E(0)\ra.
\label{eq:adia}
\end{equation}
The second equality suggests to  physically implement $|E(T)\ra$ by the following circuit:  
the first gate $\hat U (T)$ is governed by $\hat H$. The transformation $\hat S^{\dagger} (T)$ acts on the output.

{\em Adiabatic algorithms and their fast counterparts.--}
Consider now the proposition in the context of a realistic AQC, i.e. 
an ensemble of qubits described by a family of slowly-varying Hamiltonians,
\begin{equation}\label{e10}
\hat h=\Gamma(t) \sum_{i}\hat X_{i}+\hat h_P(\{\hat Z_i\}).
\end{equation}
Here, $\Gamma(t)$ is large at $t=0$, and slowly evolves towards zero at $t=T$. 
The Hamiltonian $\hat h_P(\{\hat Z_i\})$ contains the $\hat Z_i$ component of the $i$-th qubit.
The solution of a {\em hard} problem is encoded within $\hat h_P$.  
For example, Grover's search problem~\cite{Lidar} is realized with
\begin{equation}\label{e11}
\hat h_P({\hat Z_i})=\hat I-|B\ra \la B|,
\end{equation}
where $|B\ra$ is the {\em marked} state, 
and  $|B\ra \la B|$ is a function of $\hat Z_i$.  

In the D-Wave system the Hamiltonian (\ref{e10}) is given by
\begin{equation}\label{e12}
\hat h_P({\hat Z_i})=\sum_{i} h_{i}\hat Z_{i}+\sum_{ij} J_{ij}\hat Z_i \hat Z_j,
\end{equation}
with the parameters $h_i$ and $J_{ij}$.
Applying a fast magnetic field, we can enable the corresponding fast-varying Hamiltonian
\begin{equation}\label{e13}
\hat H=\gamma(t)\sum_{i}\hat X_{i}+\hat h_P(\{e^{-i\phi_{i}(t)\hat X_{i}}\hat Z_{i}e^{i\phi_{i}(t)\hat X_{i}}\}),
\end{equation}
where $\gamma(t)=\Gamma(t)+\dot{\phi}$ is a fast-varying function. 
Here, the transformation matrix is given by $\hat S(t)=\Pi_i e^{-i\phi_i(t)\hat X_i }$.
Thus, instead of evolving the system slowly under $\hat h$, the two gates  $\hat U(t)$ and
$\hat S(t)$ should be realized, Eq. (\ref{eq:adia}), allowing for a fast implementation.

{\em Built-in adiabaticity} The discussion above focuses on implementing fast dynamics to
achieve the adiabatic result, i.e. to replace slow dynamics by fast dynamics. Here we
show that the opposite is also the case, That is, fast dynamics can be shown to have a
"hidden adiabaticity".  As an example, consider a qubit under external fields, with the NMR-type Hamiltonian 
\begin{equation}\label{e5}
\hat H=\frac{\omega_0(t)}{2}\hat Z+g\left[\hat X\cos\phi(t)+\hat Y\sin\phi(t)\right].
\end{equation}
Here,  $\hat X$, $\hat Y$ and $\hat Z$ are the Pauli operators,
$\omega_0(t)$ and $\phi(t)$ potentially depend on time and are allowed to be fast-varying. 
A unitary transformation $\hat S=\exp\left[i\frac{\theta(t)-\phi(t)}{2}\hat Z\right]$ 
brings $\hat H$ into $\hat h$, 
\begin{equation}\label{e6}
\hat h=\frac{\omega_0(t)+\dot{\theta}-\dot{\phi}}{2}\hat Z+g\left[\hat X\cos\theta(t)+\hat Y\sin\theta(t)\right].
\end{equation}
We assume that $g$ is a constant and that the newly introduced time-dependent parameter $\theta(t)$, as well as
$\frac{\omega_0(t)+\dot{\theta}-\dot{\phi}}{2}$ are controlled such that they vary slowly. 
The transformation $\hat S$ thus brings the system into the adiabatic domain.
In other words, a system driven by fast-varying $\hat H$ has
built-in hidden adiabaticity characterized by $\hat h$. 
%As stated in the previous section, the existence is universal: For any given $\hat H$ there exists a corresponding adiabatic $\hat h$. 

In the particular case where $\phi(t)=\omega t$ 
and $\omega_0$, $\omega$ are constants, 
we can easily obtain the solution, that is the time evolution operator corresponding to $\hat H$,
\begin{equation}\label{e7}
\hat U=\exp\left(-i\frac{\omega \hat Z}{2} t\right)\exp\left(-i\frac{2g\hat X-\Omega \hat Z}{2}t\right)
\end{equation}
where $\Omega=\omega-\omega_0$. 
We can control parameters and realize the function $\theta(t)=\Omega t$, resulting in
\begin{equation}\label{e8}
\hat u=\exp\left(-i\frac{\Omega \hat Z}{2}t\right)\exp\left(-i\frac{2g\hat X-\Omega \hat Z}{2}t\right).
\end{equation}
The instantaneous eigenstates of $\hat h=\exp\left(-i\frac{\Omega \hat Z}{2}t\right)\hat X\exp\left(i\frac{\Omega \hat Z}{2}t\right)$ are
\begin{equation}\label{e9}
|E_{\pm}(t)\ra=\exp\left(-i\frac{\Omega \hat Z}{2} t \right)|\pm\ra.
\end{equation}
These states are proportional to the wave function $\hat u|\pm\ra$  ($\hat X|\pm\ra=\pm|\pm\ra$)
as stated by 
the adiabatic theorem for the adiabatic regime $g\gg \Omega$.
Therefore, in order to physically realize $|E_{\pm}(T)\ra$, say at $T=\pi/2\Omega$ 
when $\hat h(T)=g\hat Y$, one needs to implement two gates: $\hat U(T)$,  then $\hat S^\dagger(T)= 
\exp(i\frac{\pi \omega_0}{4\Omega} \hat Z)$.

We now come to a simple but nontrivial corollary following immediately from the Transformability proposition. 

{\em The transformability corollary at different times.--}   % DDD changed title
Let $\hat H$ (the fast Hamiltonian) be a function of the normalized or scaling time $\tau=t/T$, 
where $T$ a characteristic time of the dynamical system. Eq.~(\ref{e1}) can then be rewritten as
\begin{equation}\label{e14}
i\pa_\tau\hat U(\tau)=T\hat H(\tau)\hat U(\tau).
\end{equation}
Likewise, 
\begin{equation}\label{e15}
i\pa_\tau\hat u(\tau)=T'\hat h(\tau)\hat u(\tau),
\end{equation}
where $\tau=t'/T'$ and the latter describes a slower process so that $T<T'$, $t'(T')$ is the real time 
(characteristic time) of the Schr\"odinger equation (\ref{e15}). 
The scaling times of two equations may be identical or different. Here we set the 
same scaling time $\tau$ with the constraint $t'/T'=t/T$. 
As \red{proved} in the transformability proposition, mathematically there is at least one unitary operator $\hat S$ such that   
\begin{equation}\label{e16}
T'\hat h=\hat S^{\dagger} T\hat H \hat S-i\hat S^{\dagger}\pa_\tau{\hat S},
\end{equation}
and
\begin{equation}\label{e17}
T\hat H=\hat S T'\hat h \hat S^{\dagger}-i\hat S\pa_\tau{\hat S}^{\dagger},
\end{equation}
for a given scaling time $\tau$. 
The simplest non-trivial example is $\hat S=1$, 
such that $\hat H=\frac {T'}{T}\hat h$ and $\hat U(\tau)=\hat u(\tau)$. 
The latter equality, rewritten as $\hat U(t)=\hat u(\frac{T'}{T}t)$ with $T<T'$, 
is an exact proof that the runtime of an adiabatic quantum process can be reduced $\frac {T'}{T}$ times -- 
exact trade-off between energy and time. \red{Specifically,  Eq.~(\ref{e7}) can be rewritten as
\begin{equation}\label{e18}
\hat U(\tau)=\hat u(\tau)=\exp(-i\pi Z\tau)\exp\left(-i(T g\hat X-\pi \hat Z)\tau\right),
\end{equation}}
\red{where $\tau=t/T=t'/T'$, $gT=g'T'$ and we have set $\omega_0=0$. $U(\tau)$ ($\hat u(\tau)$) may denote a fast (adiabatic) evolution if $T'$ is in the adiabatic regime while $T$ is not in.} This result suggests a strategy of 
experimentally implementing an expedited adiabatic processes: simply enhancing the strength of the driving 
Hamiltonian to its strongest possible value.

\red{In general, the universal existence of $S$ and the equality
\begin{equation}\label{e199}
\hat u(\frac {t'}{T'})=\hat S^{\dagger}(\frac {t}{T}) \hat U(\frac {t}{T})
\end{equation}
manifests that an adiabatic quantum algorithm can always be mimicked by at most two fast gates where $T' \gg T$ is in the adiabatic regime.}

{\em Conclusion.--}Two \red{arbitrarily given} Hamiltonians within the same Hilbert space can be 
always transformed to each other via a unitary transformation.
This seemingly \red{simple but rigorous theorem} is powerful: It allows one to implement a slowly varying evolution within a fast
protocol, which is less susceptible to errors. 
We exemplified this result on a qubit system and on problems in the context of quantum adiabatic computing. 
The transformability of open quantum system Hamiltonians is left for future work. 

\acknowledgments

L.A. Wu acknowledges grant support from the Basque 
Government (Grant No. IT986-16), the Spanish MICINN (Grant No. FIS2015-67161-P).
D. S. acknowledges the Canada Research Chairs Program. We thank Professor P. Brumer for very helpful discussions.

%==================================================================

\end{document}